\documentclass[11pt,a4paper,oneside]{article}

\usepackage[natbib,style=iso-authoryear, maxcitenames=2]{biblatex}
\usepackage[nottoc]{tocbibind} 

\usepackage{hyperref}
\hypersetup{
	colorlinks=true,
	filecolor=magenta,      
	urlcolor=cyan,
	pdftitle={Real-Time Instanton Dynamics in a Triple-Well to Double-Well Homotopy},
	pdfpagemode=FullScreen,
}

\usepackage{geometry}
\usepackage[most]{tcolorbox}
\usepackage{xcolor}

\usepackage{graphicx}

\usepackage{physics}

\usepackage[scr=boondox, scrscaled=1.0]{mathalpha} 
\usepackage{amsmath,amssymb,amsfonts}
\usepackage{mathtools} 

\usepackage{chngcntr}
\counterwithin{figure}{section} 

\newcommand{\velg}[1]{\Bigl[ {#1} \Bigr]}

\newcommand{\velz}[1]{\Bigl({#1}\Bigr)}

\newcommand{\hmez}{\hspace{0.4 cm}}

\allowdisplaybreaks
\usepackage{authblk}

\title{Real-Time Instanton Dynamics in a Triple-Well to Double-Well Homotopy}

\author[1]{Vojtěch Loubal \thanks{\texttt{vojtech.loubal922@student.cuni.cz}}}
\author[1,2]{Tomáš Sýkora \thanks{\texttt{tomas.sykora@cern.ch}}}
\author[3]{Šimon Kos \thanks{\texttt{simonkos@fav.zcu.cz}}}

\affil[1]{Institute of Particle and Nuclear Physics, Faculty of Mathematics and Physics, Charles University, Prague, Czech Republic}
\affil[2]{Department of Low-Temperature Physics, Faculty of Mathematics and Physics, Charles University, Prague, Czech Republic}
\affil[3]{Department of Physics and NTIS – European Centre of Excellence, University of West Bohemia in Pilsen, Univerzitní 8, 301 00 Pilsen, Czech Republic}

\date{}

\begin{document}
\maketitle
\begin{abstract}
We show that the singularities which the analytically continued double-well instanton develops in real (Minkowski) time are not an inherent feature of real-time tunneling solutions. Rather, they can be viewed as arising from a singular limit of a one-parameter family of regular real-time solutions. Using a linear homotopy between a triple-well and a double-well potential, we construct a family of potentials, parametrized by $\mathscr{p}\in(0,1)$, whose real-time ($\alpha=0$) instanton solutions are everywhere regular and bounded. For any Wick rotation angle $\alpha\in(0,\pi/2)$, we prove that regularity is generic up to a measure-zero set. There exists a countably infinite, closed-form family of potential parameters $\{\mathscr{p}^{\mathrm{singular}}_{k}(\alpha)\}_{k\geq0}$ at which the instanton becomes singular at exactly two points of the rotated time axis, never more. In the strict double-well limit, $\mathscr{p}\to 1$ taken at $\alpha=0$, the entire family collapses onto the single point $(\mathscr{p},\alpha) = (1,0)$ at which the singular pairs of all $k$ occur simultaneously. This phenomenon produces the infinite singular comb found by Cherman and Ünsal. By solving the complexified equations of motion in terms of Weierstrass elliptic functions, we trace this comb to classical imaginary turning points receding to $\pm\mathrm{i}\infty$, giving a unified geometric and algebraic account of when, and exactly how often, real-time instantons become singular. We further show that the action of the continued instanton is independent of $\alpha$ only in its real part: its imaginary part jumps by one real-time period action at each singular angle, the $\alpha$-independence found by Cherman and Ünsal being the degenerate case of the double well, in which no such jump occurs.
\end{abstract}

\section{Introduction}

Real-time instanton solutions of quantum-mechanical systems remain a crucial yet challenging area of instanton analysis \citep{Nishimura2023jhep}, \citep{Nishimura2023}, \citep{Levkov2005a}, \citep{Tanizaki2014}. Understanding their dynamics would illuminate  real-time scattering phenomena, tunneling dynamics and the foundational structure of the real-time path integral.

A valuable approach was taken by Cherman and Ünsal \citep{Cherman2014}. They proceeded by analytically continuing the double-well kink solution using the complex Wick rotation by an arbitrary angle. As they showed, the kink solution develops periodic singularities in real time. They also showed that the action of the continued kink does not depend on the rotation angle $\alpha$, and conjectured that this solution has the minimal non-zero action for every $\alpha$. We refer to this as the Cherman-Ünsal conjecture. In later works, it was concluded using resurgence theory and Picard-Lefschetz theory that singular solutions are equally important to the path integral as regular ones \citep{Behtash2015}. In a related direction, Ai, Garbrecht and Tamarit \citep{Ai2019a} recovered the one-loop Callan-Coleman decay rate of the false vacuum directly from the analytically continued instanton, providing an independent real-time derivation of a result previously obtained via the Euclidean path integral, thereby further expanding the Picard-Lefschetz framework.

In this work, we demonstrate that the singular nature of the real-time double-well instanton is not an inherent feature of real-time tunneling. By constructing a linear homotopy between a symmetric triple-well and a double-well potential, we analyze a family of systems possessing fully regular real-time instanton solutions. The singular double-well solution can then be understood as a degenerate limiting case within this family.

We show that as the potential deforms into the double-well limit, imaginary turning points in the complex plane move to infinity, giving rise to real-time divergences. Furthermore, we establish that for each $\alpha \in (0,\pi/2)$ there is a countable, closed-form family $\{\mathscr{p}^{\mathrm{singular}}_{k}\}_{k \geq 0}$, where each member produces exactly two singularities on the rotated time axis and never more. The infinite comb is then a property of the single point ($\mathscr{p},\alpha) = (1,0)$. Moreover, the singular angles are exactly where the action of the continued instanton jumps: its real part is independent of $\alpha$ throughout the family, while its imaginary part changes by one real-time period action at each singular angle.

Finally, by solving the complexified equations of motion in terms of Weierstrass elliptic functions, as established for the double well (see below), we provide an algebraic and geometric classification of these instantons, mapping their singularities directly to the degenerate, singly periodic limit of the doubly periodic lattice of the complex plane. 

Complex classical solutions of the double well have a long history. Richard and Rouet classified all real and complex saddle points of the double well at finite Euclidean period as Weierstrass elliptic functions \citep{Richard1981a,Richard1981b}. Such methods were soon applied beyond the quartic: for the octic double well, energy levels and splittings were obtained semi-classically where dilute-gas instanton techniques do not apply \citep{Mottola1982}, and the classification was later rediscovered by Nekrasov, who labelled the saddles by winding numbers around the cycles of the spectral curve \citep{Nekrasov2018}. A selection rule for the saddles based on deforming the elapsed-time contour was proposed by Carlitz and Nicole \citep{Carlitz1985}. In real time, Tanizaki and Koike identified all complex saddle points of the real-time double-well action, labelled by an integer lattice, showed that real-time tunneling is described by them, and identified the determination of their thimble intersection numbers as an open problem \citep{Tanizaki2014}. Behtash et al.\ introduced critical points at infinity, complex saddles pushed to infinity as $T\to\infty$ while their action stays finite \citep{Behtash2018}, and bion and instanton contributions in triple-well and multi-well potentials were analysed by Dunne, Sulejmanpasic and Ünsal \citep{Dunne2020}. Most recently, Dersy and Schwartz used the exact finite-$T$ Weierstrass saddles of the double well to organize its complete resurgent structure through Picard-Lefschetz integrals over the quasi-zero modes, showing in particular that only real saddles contribute to the Euclidean partition function, while the complex ones govern its Stokes structure \citep{Dersy2026a,Dersy2026b}. Feldbrugge, Jow and Pen showed that complex classical paths in the real-time path integral can cross singularities, beyond which they no longer solve the boundary-value problem and must be analytically continued, the action acquiring a discrete correction that is needed to reproduce the tunneling rate \citep{Feldbrugge2023,Feldbrugge2025}. Steingasser and Kaiser described tunneling in the double well in strictly real time, regularizing the Hamiltonian instead of rotating the time contour \citep{Steingasser2025}.

All of these works concern a fixed potential. In the present work we instead vary the potential along a homotopy and ask how the singularities of the continued instanton on the rotated time axis depend on it. In the language of \citet{Dersy2026a,Dersy2026b}, our $C=0$ solutions sit at the $T\to\infty$ corner of the finite-$T$ elliptic curve, where the classical energy vanishes, one period diverges and the torus pinches to a cylinder. We probe this corner not by winding at finite $T$ but by rotating the time contour. What is new here is therefore not the Weierstrass representation itself, which we carry over to the sextic family, but the regularizing homotopy, the exact classification of singularities on the rotated time axis it yields, the identification of their mechanism, and an exactly solvable instance of the singularity crossings of \citet{Feldbrugge2023,Feldbrugge2025} in the setting of Cherman and Ünsal (Sec.~\ref{sec:action}).

The paper is organized as follows. In Section~\ref{sec:homotopy} we construct
the homotopy family and identify the critical parameter $\mathscr{p}_c$ at which the central well appears. In Section~\ref{sec:euclidean} we present the exact Euclidean
instanton, its splitting into two half-instantons, and the closed-form action corresponding to this configuration. In Section~\ref{sec:continuation} we perform the complex Wick rotation and discuss 
the behaviour of singularities within the homotopy family. In Section~\ref{sec:weierstrass} we derive the
Weierstrass representation of the general complex solution and identify the
geometric origin of the singularities, and compute the action of the continued instanton. In Section~\ref{sec:discussion} we summarize and discuss extensions.

\section{Homotopy potential}\label{sec:homotopy}
Consider a symmetric $n$-well potential defined by
\begin{equation}
    V_n(z) = P_{n}(z^2),
\end{equation}
where $P_n$ is a polynomial of degree $n$ with a positive leading coefficient. Energy conservation
\begin{equation}\label{def_EOM}
    (\dot{z})^2 = 2(C-V_n(z)) = 2(C-P_n(z^2)),
\end{equation}
where $C \in \mathbb{C}$ is the complex energy (an integration constant), and the dot denotes differentiation with respect to the time
$t$, can be simplified using the $Z_2$ symmetry $z\to-z$ of the system. Making a substitution $w = z^2$ yields
\begin{equation}
    (\dot{w})^2 = 8w(C-P_n(w)) = R_{n+1}(w).
\end{equation}
Substituting $w=z^2$ reduces the first integral (\ref{def_EOM}) to a polynomial of degree $n+1$ in $w$ possessing a root at $w = 0$ by construction. 

A class of potentials can then be studied via the linear homotopy
\begin{equation}\label{eq:homotopy}
    V_{\mathscr{p}}(z) = (1-\mathscr{p})P_{n}(z^2) + \mathscr{p} Q_{l+1}(z^2), \qquad n = 2l+1,\ l \in \mathbb{Z}_{\geq 1},
\end{equation}
where $P_n$ and $Q_{l+1}$ are polynomials of degrees $n$ and $l+1$ in $w=z^2$. Both endpoint potentials are thereby $\mathbb{Z}_2$-symmetric, no half-integer powers of $w$ appear, and the above construction implies the first integral
\begin{equation}
    (\dot{w})^2 = 8w(C-(1-\mathscr{p})P_n(w)-\mathscr{p}Q_{l+1}(w)) = R_{n+1}(w),
\end{equation}
which is exactly solvable by a hyperelliptic function of genus $g = \lfloor \frac{n}{2} \rfloor = l$.

We consider the simplest case using a linear homotopy between a symmetric triple well potential $V_3(z) = \lambda z^2(z^2-\eta^2)^2$ and a symmetric double well potential $V_2(z) = \lambda (z^2-\eta^2)^2$ given by:
\begin{equation}\label{def:Homotopy_potential}
    V_{\mathscr{p}}(z) = \lambda (z^2-\eta^2)^2 \velg{(1-\mathscr{p})z^2 + \xi^2 \mathscr{p}},
\end{equation}
where $\lambda$ acts as a coupling constant, $\eta$ determines the position of two lateral minima and $\xi$ is a geometric constant needed for the system to be dimensionally consistent. From the above discussion, we expect the general solution to (\ref{def:Homotopy_potential}) to have a form of an elliptic (genus-one) function. 

The homotopy (\ref{eq:homotopy}) interpolates between an $n$-well and an $(l+1)$-well. These well-counts are consecutive integers only for $l = 1$ from which it follows that $n=3$. The triple-well/double-well system (\ref{def:Homotopy_potential}) is therefore the unique member of this family of constructions connecting potentials with consecutive numbers of wells. For $l \geq 2$, the homotopy necessarily skips over intermediate well-counts.

We emphasize that the family (\ref{def:Homotopy_potential}) is not an arbitrary choice of triple-well/double-well interpolation but it is selected precisely because the associated Bogomol'nyi-Prasad-Sommerfield (BPS) quadrature is elementary and explicitly invertible. This controllability is what makes an exact treatment of the analytic continuation possible in closed form throughout this paper, and should be read as a deliberate design choice rather than an incidental property of a generic homotopy.

\begin{figure}[h!]
    \centering
    \includegraphics[width=0.9\linewidth]{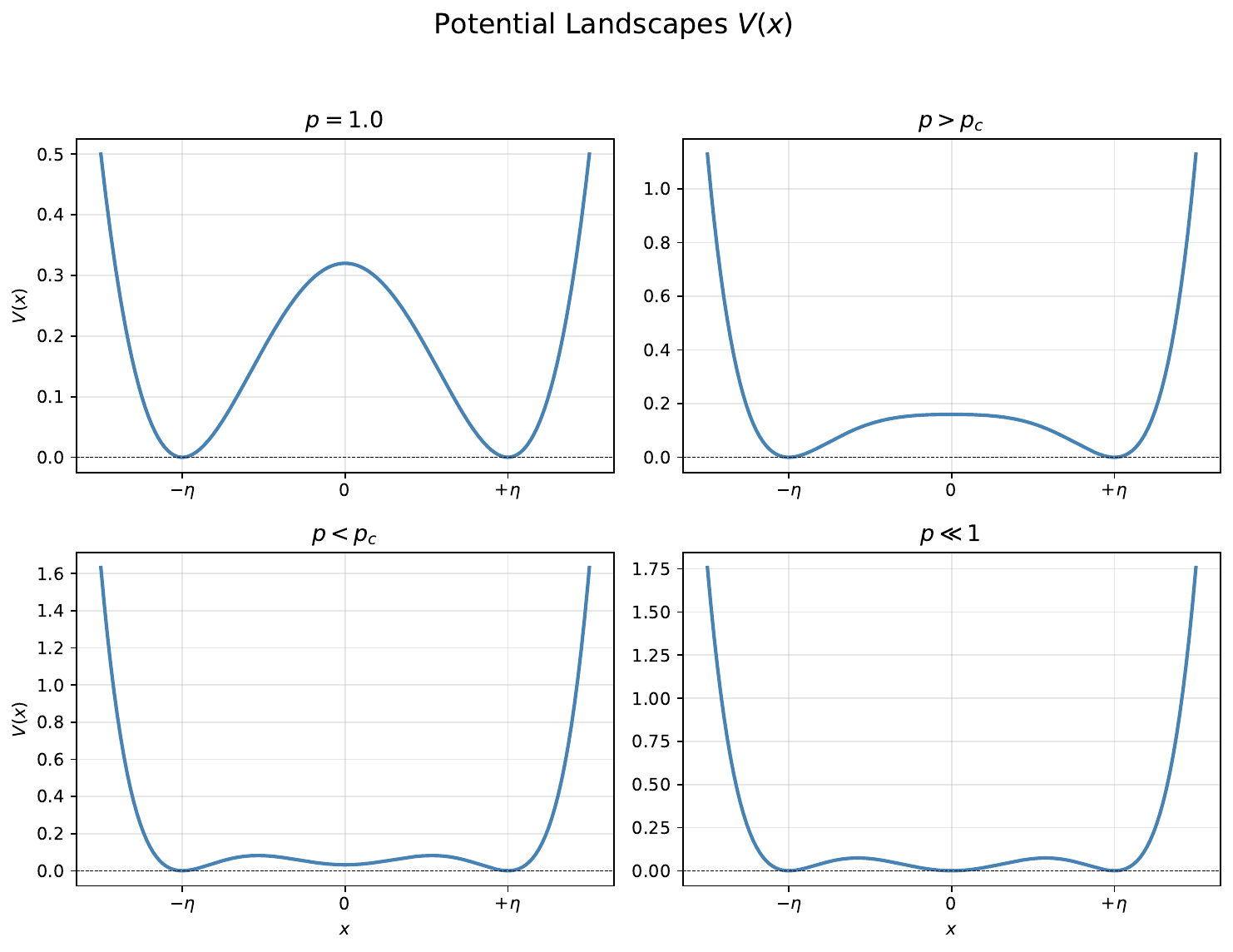}
    \caption[Shape of the potential $V_\mathscr{p}(x)$ as a function of the homotopy parameter $\mathscr{p}$.]{Shape of the potential $V_\mathscr{p}(x)$ as a function of the homotopy parameter $\mathscr{p}$. At $\mathscr{p} = 1$, the system exhibits the classic double-well structure with vacua located at $\pm \eta$ and a primary barrier at the origin. As $\mathscr{p}$ decreases through the regime $\mathscr{p} > \mathscr{p}_c$, the central barrier is suppressed until the critical value $\mathscr{p}_c$ is reached, at which point the curvature at the origin changes sign. In the $\mathscr{p} < \mathscr{p}_c$ regime, a secondary local minimum begins to emerge at $x = 0$, transforming the central barrier into two distinct peaks. Finally, in the limit $\mathscr{p} \ll 1$, the system matures into a full triple-well potential. Parameters used: $\lambda = 0.5, \eta = 1, \xi = 0.8, \mathscr{p} \in \{1.0, 0.5, 0.1, 0.00001\}$, for which $\mathscr{p}_c \approx 0.44$.}
    \label{fig:placeholder}
\end{figure}

To study the instanton behaviour, we first evaluate the fundamental frequencies given by second derivatives of the potential (\ref{def:Homotopy_potential}) at the minima
\begin{align}
    \omega^2& \coloneqq  V''(\pm \eta)   = 8 \lambda \eta^2 [(1-\mathscr{p})\eta^2 + \xi^2 \mathscr{p}], \\
    \omega_M^2 & \coloneqq V''(0) =   2\lambda\eta^2[(1-\mathscr{p})\eta^2 -2\xi^2 \mathscr{p}].
\end{align}

 A critical value $\mathscr{p}_c$ exists where the central well flattens ($\omega_M^2 = 0$)
\begin{equation}\label{eq:critical_p}
    \mathscr{p}_c = \frac{\eta^2}{2\xi^2+\eta^2} = \frac{(\eta / \xi)^2}{2+(\eta/\xi)^2},
\end{equation}
 and we transition between a double-well and a triple-well. The value of $\mathscr{p}_c$ is determined purely by the geometry of the system included in the ratio $\eta/\xi$ and is independent of the coupling constant $\lambda$.

\section{Euclidean instanton solutions}\label{sec:euclidean}
The instanton solution follows the BPS equation \citep{Coleman1979}
\begin{equation}\label{def:BPS_equation}
    \dot{X}_\mathscr{p} = \pm \sqrt{2V_\mathscr{p}(X)},
\end{equation}
where the dot now denotes the derivative with respect to the Euclidean
time $\tau$, and is given by
\begin{equation}\label{eq:instanton_solution}
    X_\mathscr{p}^{\pm}(\tau) = \pm \frac{\eta \sinh\left(\frac{\omega(\tau-\tau_c)}{2}\right)}{\sqrt{\cosh^2\left(\frac{\omega(\tau-\tau_c)}{2}\right)+\frac{1-\mathscr{p}}{\mathscr{p}}\frac{\eta^2}{\xi^2}}}. 
\end{equation}

\begin{figure}[h!]
    \centering
    \includegraphics[width=0.9\linewidth]{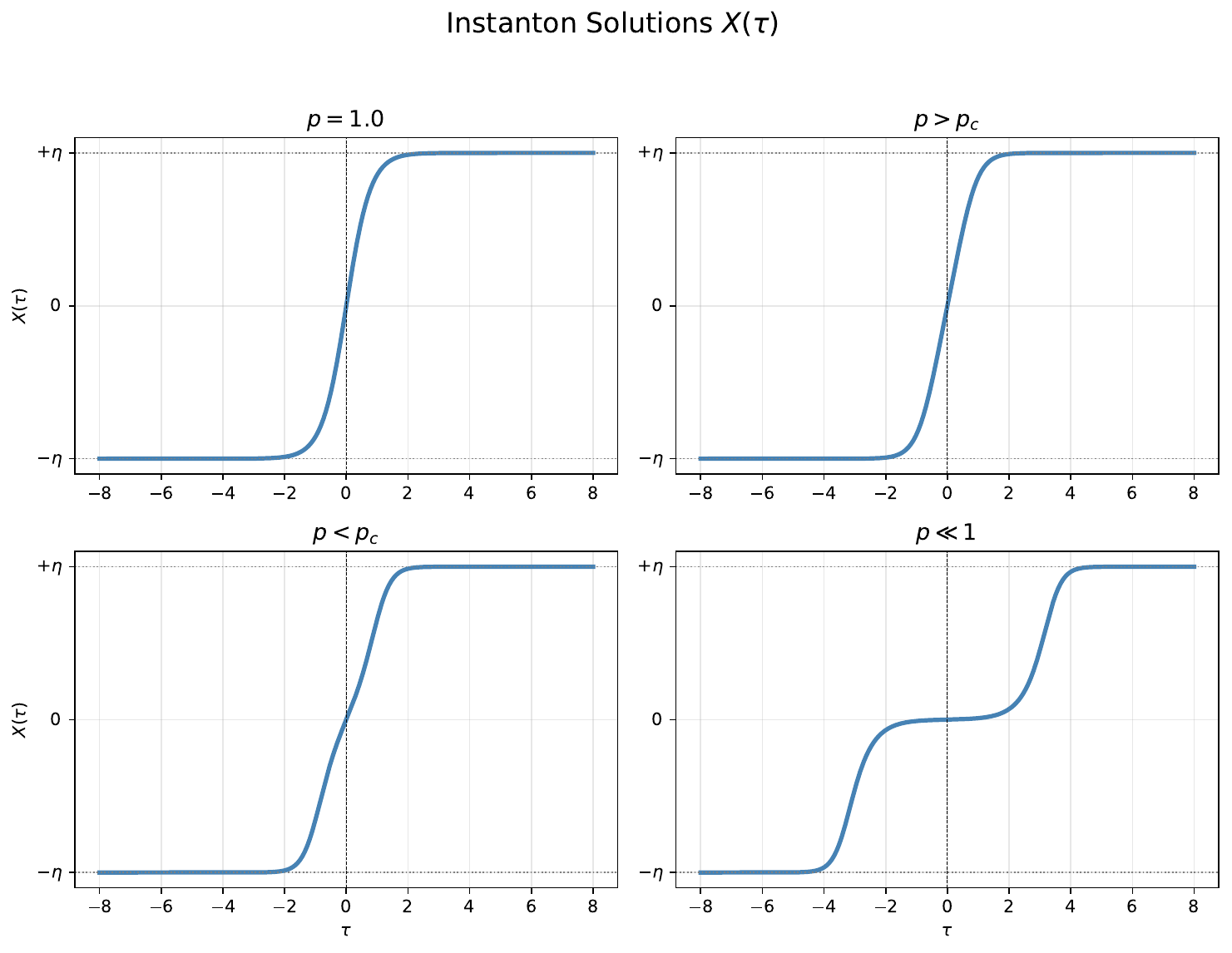}
    \caption[Evolution of the Euclidean instanton solutions $X_\mathscr{p}(\tau)$ for distinct values of the parameter $\mathscr{p}$.]{Evolution of the Euclidean instanton solutions $X_\mathscr{p}(\tau)$ for distinct values of the parameter $\mathscr{p}$. In the double-well limit ($\mathscr{p} = 1$), the solution is a standard kink interpolating directly between the vacua at $-\eta$ and $+\eta$. As $\mathscr{p}$ decreases toward the critical value $\mathscr{p}_c$, the trajectory remains a single transition. In the triple-well regime ($\mathscr{p} < \mathscr{p}_c$), the solution begins to exhibit a double-transition structure, which becomes most prominent in the $\mathscr{p} \ll 1$ limit. In this regime, the particle tunnels from $-\eta$ to the intermediate minimum at $0$, dwells there and subsequently tunnels to the final vacuum at $+\eta$. Parameters used: $\lambda = 0.5, \eta = 1, \xi = 0.8, \tau_c = 0, \mathscr{p} \in \{1.0, 0.5, 0.1, 0.00001\}$.}
    \label{fig:instantons}
\end{figure}
The positive solution is an instanton and the negative solution is an anti-instanton.

By defining the dimensionless argument
\begin{equation}\label{def:argument}
    u = \frac{\omega (\tau-\tau_c)}{2},
\end{equation}    
and the dimensionless ratio
\begin{equation}\label{def:b}
    b^2 = \velz{\frac{\eta}{\xi}}^2\frac{1-\mathscr{p}}{\mathscr{p}},
\end{equation}  
the solution (\ref{eq:instanton_solution}) can be written as 
\begin{equation}\label{def:instanton_solution}
X^{\pm}_\mathscr{p}(\tau) = \pm \frac{\eta \sinh(u)}{\sqrt{\cosh^2(u) + b^2}}.
\end{equation}

In terms of $b$, the potential (\ref{def:Homotopy_potential}) takes the one-parameter shape
\begin{equation}
V_\mathscr{p}(z)=\Lambda\,(z^2-\eta^2)^2\Bigl(1+\frac{b^2}{\eta^2}z^2\Bigr),
\qquad \Lambda\equiv\lambda\xi^2\mathscr{p},
\label{eq:oneparam}
\end{equation}
with $\omega^2=8\Lambda\eta^2(1+b^2)$. Up to the overall scale $\Lambda$ and
the length $\eta$, the family is thus parametrized by $b$ alone. The
double-well limit $\mathscr{p}\to1$ corresponds to $b\to0$, while the triple-well limit
$\mathscr{p}\to0$ corresponds to $b\to\infty$ taken jointly with $\Lambda\to0$.

A key phenomenon occurs for values of the parameter $\mathscr{p} \leq \mathscr{p}_c$. The instanton solution splits into two half-instantons which corresponds to tunneling through the middle well (see Fig.~\ref{fig:instantons}). We characterize these half-instantons by their instanton centers which we calculate using the inflection points of the solution (\ref{def:instanton_solution}). Unlike the two instanton valley solution \citep{Balitsky1986}, the instanton centers of the two half-instantons are determined by the system because the instanton (\ref{def:instanton_solution}) is an exact solution to the BPS equation (\ref{def:BPS_equation}).

For $\mathscr{p} \leq \mathscr{p}_c$ there exist three inflection points $\{\tau'_i\}_{i = 0,1,2}$ given by
\begin{equation}
    \tau'_0 = \tau_c,
\end{equation}
and
\begin{equation}\label{eq:inflection_points}
    \tau'_{1,2}(\mathscr{p}) = \tau_c \pm \frac{2}{\omega} \mathrm{arcosh}\velz{\frac{b(\mathscr{p})}{\sqrt{2}}}.
\end{equation}
We define the half-instanton separation $R$ which traces a curve in the functional space of solutions by
\begin{equation}\label{eq:R}
    R(\mathscr{p}) \coloneqq \tau'_2 - \tau'_1 = \frac{4}{\omega}\mathrm{arcosh}\velz{\frac{b(\mathscr{p})}{\sqrt{2}}}.
\end{equation}

The value of the classical Euclidean action $S_{\mathrm{E}}^0$ for the instanton configuration (\ref{def:instanton_solution}) results in
\begin{equation}\label{eq:action}
    S_{\mathrm{E}}^0(b) = \sqrt{2\lambda}\eta^4\xi \velg{\frac{(2b^2-1)\sqrt{1+b^2}}{4b^2\sqrt{\eta^2+\xi^2b^2}}+\frac{4b^2+1}{4b^3\sqrt{\eta^2+\xi^2b^2}} \mathrm{artanh} \sqrt{\frac{b^2}{1+b^2}}}.
\end{equation}

Since $\sqrt{\eta^2+\xi^2b^2}=\eta/\sqrt{\mathscr{p}}$, Eq.~(\ref{eq:action}) can equivalently be written, with $\Lambda$ of Eq.~(\ref{eq:oneparam}), with the $b$-dependence isolated in a single dimensionless bracket,
\begin{equation}
S_{\mathrm{E}}^0(b)=\sqrt{2\Lambda}\;\eta^3\Biggl[
\frac{(2b^2-1)\sqrt{1+b^2}}{4b^2}
+\frac{4b^2+1}{4b^3}\,\mathrm{artanh}\sqrt{\frac{b^2}{1+b^2}}\Biggr].
\end{equation}
Moreover, the transcendental factor satisfies the exact identity
\begin{equation}
\mathrm{artanh}\sqrt{\frac{b^2}{1+b^2}}
=\log\bigl(b+\sqrt{1+b^2}\bigr)=\mathrm{arsinh}(b)\equiv\theta,
\label{eq:theta}
\end{equation}
and the same quantity $\theta$ will reappear both in the singularity condition
of Sec.~\ref{sec:continuation} and in the pole geometry of
Sec.~\ref{sec:weierstrass}.

In the limit as $\mathscr{p} \to 1, b \to 0$ and we find that
\begin{equation}
    S_\mathrm{E}^0 \xrightarrow{\mathscr{p} \to 1} \frac{4}{3}\sqrt{2\lambda} \eta^3 \xi = \frac{4}{3}\sqrt{2\Lambda} \eta^3,
\end{equation}
where $\Lambda = \lambda \xi^2$ at $\mathscr{p}=1$, which agrees with the double-well action for the potential $V_{\mathscr{p}=1}(x) = \Lambda (x^2-\eta^2)^2$. As $\mathscr{p} \to 0, b \to \infty$ the action approaches
\begin{equation}\label{eq:triplewell_action}
    S_\mathrm{E}^0 \xrightarrow{b \to \infty}  \sqrt{2\lambda} \frac{\eta^4}{2} \Big( 1 + \frac{4 \log(2b)-(\eta/\xi)^2}{2 b^2}\Big).
\end{equation}
which reduces to twice the triple-well action as expected. The scale of
the corrections is set by the half-instanton separation. From
Eq.~\eqref{eq:R}, $e^{-\omega R/2}\to 1/(2b^2)$ at large $b$, and since $\omega/2\to\omega_M$ in this limit, the overall
$1/b^2$ factor is precisely the exponential kink-kink interaction scale $e^{-\omega_M R}$ of the
multi-instanton expansion \citep{ZinnJustin1981}. The logarithm
$\log(2b)\simeq \omega R/4$ reflects the fact that for $\mathscr{p}>0$ the central minimum is lifted to $V_\mathscr{p}(0)=\lambda\eta^4\xi^2\mathscr{p}=\Lambda\eta^4>0$. To first order, such a lift changes the action by the integral of the potential shift along the path, and between the two half-kinks the path spends the time $R$ near $z=0$. The plateau therefore contributes $V_\mathscr{p}(0)\,R\simeq(4/\omega)\,V_\mathscr{p}(0)\log(2b)\simeq\sqrt{2\lambda}\,\eta^4\log(2b)/b^2$, which is exactly the logarithmic term of Eq.~(\ref{eq:triplewell_action}). The remaining corrections are of order $1/b^2$ without the logarithm. This term grows linearly with the separation $R$, but its coefficient $V_\mathscr{p}(0)\simeq\lambda\eta^6/b^2$ is itself of order $1/b^2$ and vanishes in the strict triple-well limit. That $V_\mathscr{p}(0)$ and $e^{-\omega_M R}$ are of the same order is not accidental: since the BPS solution saturates the Bogomol'nyi bound and is therefore the minimum of the action in its sector, the separation $R$ is fixed by the balance between this linear cost and the short-range repulsion of the two half-kinks.

\section{Analytical continuation and the structure of singularities}\label{sec:continuation}
In their work \citep{Cherman2014}, Cherman and Ünsal considered the analytical continuation of the instanton solution using the complex Wick rotation
\begin{equation}\label{def:complex_time}
    t = e^{-\mathrm{i}\alpha} \tau, \hmez \tau \in \mathbb{R}.
\end{equation}
The complex time $t$ represents a line in the complex plane passing through the origin $t=0$. For $\alpha = 0$ we describe real time and for $\alpha = \pi/2$ we recover Euclidean time.

We apply this transformation to our solution (\ref{def:instanton_solution}) yielding
\begin{equation}\label{eq:complex_time_trajectory}
    X_\mathscr{p}^{\alpha}(\tau) = \frac{\eta \sinh(\frac{\omega}{2}(\tau-\tau_c)e^{i(\pi/2-\alpha)})}{\sqrt{\cosh^2(\frac{\omega}{2}(\tau-\tau_c)e^{i(\pi/2-\alpha)})+b^2}},
\end{equation}
the solution can be written as in (\ref{def:instanton_solution})
\begin{equation}\label{def:complex_time_trajectory}
    X_\mathscr{p}^{\alpha} (\tilde{u}) = \frac{\eta \sinh(\tilde{u})}{\sqrt{\cosh^2(\tilde{u})+b^2}}.
\end{equation}
where we define the complex argument 
\begin{equation}
    \tilde{u} = ue^{\mathrm{i}(\pi /2- \alpha)},
\end{equation}
with $u$ being the real argument as in Eq.~(\ref{def:argument}). Here and below we continue the instanton branch $X^{+}_\mathscr{p}$ of Eq.~(\ref{def:instanton_solution}) and the anti-instanton branch follows by the overall sign flip $X\to-X$. For a highly detailed visual representation of these trajectories and their limiting singular behaviors, the reader is referred to the full-page plots in Appendix~\ref{AppendixB}, where we show the complex trajectory in the complex plane (Fig.~\ref{Huge_phase_space}) and the time evolution of both real and imaginary parts of the analytically continued solution (Fig.~\ref{Re-Im_evolution}).

As we saw in Eq.~(\ref{eq:inflection_points}), in \emph{Euclidean} time the inflection point $\tau'_0=\tau_c$ exists for all $b$, while $\tau'_{1,2}$ exist only for $b \geq \sqrt{2}$ or $\mathscr{p} \leq \mathscr{p}_c$. In purely \emph{real} time, the inflection points of Eq.~(\ref{eq:complex_time_trajectory}) form two
infinite families,
\begin{equation}
\tau'_{0;k}=\tau_c + \frac{2\pi k}{\omega}
\qquad \text{and} \qquad
\tau'_{1,2;k}=\tau_c\pm\frac{2}{\omega}\arccos\bigl(b/\sqrt2\bigr)
+\frac{2\pi k}{\omega},\qquad k\in\mathbb{Z}.
\end{equation}
The first family exists for all $b$. The second exists only for $b\le\sqrt2$,
i.e.\ $\mathscr{p} \ge \mathscr{p}_c$, mirroring the Euclidean condition $b\ge\sqrt2$ of
Eq.~(\ref{eq:inflection_points}). Both conditions descend from the same algebraic
equation $\cosh^2\tilde{u}=b^2/2$. Both families recur with period $2\pi/\omega$ in $\tau$ which is half
the $4\pi/\omega$ period of the solution itself because $X$ is
antiperiodic under $u\to u+\pi$, $X(u+\pi)=-X(u)$, and $X$ and $-X$ share
their inflection points.

Contrary to the Euclidean time solution (\ref{def:instanton_solution}), singularities start to appear once we introduce the complex Wick rotation (\ref{def:complex_time}). In Euclidean time, the denominator of (\ref{def:instanton_solution}) is always nonzero. In complex time, this is no longer the case. Singularities occur wherever
\begin{equation}\label{def:singular_condition}
    \cosh^2(\tilde u) + b^2 = 0 \quad\Longleftrightarrow\quad \cosh(\tilde u) = \pm \mathrm{i} b.
\end{equation}

Since $\tilde u = u\sin\alpha + \mathrm{i}u\cos\alpha$ and $\cosh(u\sin\alpha)$ never vanishes, Eq.~(\ref{def:singular_condition}) is equivalent to the system of equations
\begin{align}
    \sinh(u\sin\alpha) = \pm b, \label{def:system_eq1}\\
    \cos(u\cos\alpha) = 0. \label{def:system_eq}
\end{align}
Solving the system of equations (\ref{def:system_eq1})--(\ref{def:system_eq}) gives the single compatibility condition for $k \in \mathbb{Z}$ that produces singular solutions
\begin{equation}\label{eq:compatibility}
    k+\frac{1}{2} = \pm\,\kappa(\alpha,b), \qquad \kappa(\alpha,b) \coloneqq \frac{\mathrm{arsinh}(b)}{\pi\tan(\alpha)}  = \frac{\theta}{\pi \tan(\alpha)}> 0.
\end{equation}

For fixed $\alpha \neq 0$ and $b>0$, Eq.(\ref{eq:compatibility}) admits a solution if and only if $\kappa(\alpha,b)$ equals one of the positive half-integers $\{\tfrac12,\tfrac32,\tfrac52,\dots\}$ which forms a discrete, measure-zero condition on $(\alpha,b)$. When it holds, say $\kappa(\alpha,b) = k_0+\tfrac12$ for a unique $k_0 \in \mathbb{Z}_{\geq0}$, the only integers satisfying (\ref{eq:compatibility}) are $k=k_0$ and $k=-1-k_0$ and no others, since $\kappa(\alpha,b)$ is a single fixed number. These correspond to $u = \pm(2k_0+1)\pi/(2\cos\alpha)$, i.e. two singular values
\begin{equation}\label{def:tau_singular}
    \tau_{\text{singular}} = \tau_c \pm \frac{\pi}{\omega\cos(\alpha)}(2k_0+1),
\end{equation}
symmetric around $\tau_c$. For a generic pair $(\alpha,\mathscr{p})$ the real-$\tau$ solution has no singularities at all. When it is tuned to one of these discrete values it has exactly two.

Eq.~(\ref{eq:compatibility}) can equivalently be read as fixing the homotopy parameter. Inverting $\kappa(\alpha,b)=k+\tfrac12$ for $\mathscr{p}$ gives 
\begin{equation}\label{def:p_singular}
    \mathscr{p}^{\mathrm{singular}}_{k}(\alpha) = \frac{\eta^2}{\eta^2 + \xi^2 \sinh^2\!\big(\pi \tan(\alpha)(k+\tfrac12)\big)}, \qquad k = 0,1,2,\dots
\end{equation}
Because $\mathrm{arsinh}(b)$ is a continuous, strictly increasing bijection from $(0,\infty)$ to $(0,\infty)$, for every fixed $\alpha \in (0,\pi/2)$ and every $k \geq 0$ there exists a unique $b_k(\alpha)$, and hence a unique $\mathscr{p}^{\mathrm{singular}}_{k}(\alpha) \in (0,1)$, solving Eq.~(\ref{eq:compatibility}). Only $k \geq 0$ need be listed. The reflection $k \to -1-k$ leaves the $\sinh^2$ term in Eq. (\ref{def:p_singular}) invariant, so negative $k$ reproduce the same potential (\ref{def:Homotopy_potential}), in accordance with each potential having two singularities.

It then follows that for any $\alpha \in (0,\pi/2)$ there is a countably infinite family of potentials $\{V_{\mathscr{p}^{\mathrm{singular}}_{k}(\alpha)}\}_{k\geq0}$, each producing exactly the singular pair (\ref{def:tau_singular}) at the corresponding $k$. As $k\to\infty$, $\mathscr{p}^{\mathrm{singular}}_{k}(\alpha) \to 0$, approaching the triple-well limit.

This construction degenerates in real time ($\alpha = 0$) as $\tan\alpha=0$ sends $\kappa(\alpha,b)\to\infty$ for any $b>0$. No finite-$b$ potential thus produces a real-time singularity at exactly $\alpha=0$, consistent with the bound (\ref{eq:bound}) below. The sole exception is the strict double-well limit $b=0$ ($\mathscr{p}=1$), where the compatibility condition (\ref{eq:compatibility}) collapses to $\cos(u)=0$, satisfied by every integer $k$ simultaneously. This is the infinite Dirac comb of Cherman and Ünsal \citep{Cherman2014}, a phenomenon of the $b=0,\,\alpha=0$ point specifically, not a limit approached by $\alpha \to 0$ at fixed generic $\mathscr{p}$.

We further expand on the physical significance of these divergences by considering the real-time solution
\begin{equation}
    X_\mathscr{p}^{\alpha = 0} = \mathrm{i} \frac{\eta \sin(u)}{\sqrt{\cos^2(u)+b^2}}.
\label{eq:realtime}
\end{equation}
For any $b > 0$, this function remains bounded.
\begin{equation}\label{eq:bound}
X_\mathscr{p}^{\alpha=0} \in \velg{-\mathrm{i}\frac{\eta}{b}, \mathrm{i}     \frac{\eta}{b}}.
\end{equation}

Interestingly, part of the singularity condition survives in a well-defined way as $\alpha \to 0$, independently of whether $b$ happens to be tuned to satisfy (\ref{eq:compatibility}). The real part of $\cosh(\tilde u)$ can vanish only where
\begin{equation}
    \cos(u\cos\alpha) = 0 \quad\Longleftrightarrow\quad u = \frac{(2k+1)\pi}{2\cos\alpha},
\end{equation}
since $\cosh(u\sin\alpha)\geq 1$ never vanishes. This locus is exactly the time coordinate appearing in (\ref{def:tau_singular}), and unlike the full compatibility condition (\ref{eq:compatibility}) it does not depend on $b$. It is a necessary, but not by itself sufficient, condition for a singularity, fixing where one could occur before the value of $b$ is taken into account at all. As $\alpha \to 0$, $\cos(u\cos\alpha) \to \cos(u)$, and at $u=\pi(2k+1)/2$ we have $\cos(u)=0$, so from Eq.~(\ref{eq:bound}) the position reaches exactly its extremal value $X=\pm\mathrm{i}\eta/b$, while the velocity
\begin{equation}\label{eq:imag_velocity}
\left( \frac{dX}{d\tau} \right)_{\mathscr{p}}^{\alpha = 0} = \mathrm{i} \frac{\omega \eta(1+b^2)}{2} \frac{\cos(u)}{(\cos^2(u)+b^2)^{3/2}},
\end{equation}
vanishes there, since $\cos(u)$ is in the numerator. So rather than becoming ill-defined, the candidate locus converges exactly onto the turning points $\tau_{\text{turn}} = \tau_c + \frac{\pi}{\omega}(1+2k)$ of the always-bounded, always-regular $\alpha=0$ trajectory, where the particle momentarily comes to rest before reversing direction for any $b>0$, coinciding precisely with (\ref{def:tau_singular}) evaluated at $\alpha=0$.

We also see that for any finite $b \in (0,\infty)$, the real-time solution (\ref{eq:complex_time_trajectory}) is closely related to the finite-energy periodic instantons, which govern tunneling at finite temperature, found for the double-well and sine-Gordon potentials by Liang and Müller-Kirsten \citep{Liang1992}, and by Khlebnikov, Rubakov and Tinyakov \citep{Khlebnikov1991}, using the same elliptic-function machinery. The relation is made precise in Sec.~\ref{sec:weierstrass}: at $C\neq0$ the parameter $m$ of the underlying Jacobi representation is genuinely elliptic, $m\in(0,1)$, and the corresponding finite-energy periodic orbits are the periodic instantons of \citet{Liang1992} and \citet{Khlebnikov1991} (in Euclidean time at finite period, the exact saddles used by \citet{Dersy2026a}). The tunneling boundary conditions of the present problem force $C=0$, pinching the parameter to its degenerate value, so the periodic solutions found here are the degenerate limit of the periodic-instanton family rather than generic members of it. Its period is $\frac{4\pi}{\omega}$, and the Minkowski action over one period is
\begin{equation}
    S_{\text{period}} = - \frac{\pi \eta^2 \omega}{8b^3\sqrt{1+b^2}}(1+4b^2).
\label{eq:Speriod}
\end{equation}
Because $S_{\text{period}}$ is real and negative, the corresponding weight $e^{\mathrm{i}S_{\text{period}}}$ is a pure phase. These saddles are unsuppressed and purely oscillatory, rather than exponentially suppressed as in the Euclidean case. The same quantity reappears in Sec.~\ref{sec:action} as the jump of the continued-instanton action at each singular angle.

\section{Origin of the singularities and the Weierstrassian representation}\label{sec:weierstrass}
In quantum mechanics, Picard-Lefschetz theory provides a decomposition of a path-integral measure into a sum of convergent integrals over cycles in the complexified trajectory space which are called Lefschetz thimbles. The thimbles are attached to stationary points of the Minkowski action \citep{Tanizaki2014}. A full treatment would require constructing these thimbles explicitly and computing the intersection numbers $n_\sigma$
that dictate which saddles actually contribute, together with the one-loop fluctuation determinant around the periodic real-time solutions of Section~\ref{sec:continuation}. We leave this analysis, and the resulting statement about which of our singular and periodic saddles survive in the contour decomposition, to future work. Here we restrict ourselves to the more limited question of classifying the classical complex solutions themselves. To describe the most general solutions, we begin by complexifying the position $z=x+\mathrm{i}y$. The conservation of complex energy $C$ leads to the first-order differential equation
\begin{equation}\label{def:complex_EOM}
    (\dot{z})^2 = 2[C-V_{\mathscr{p}}(z)], 
\end{equation}
where $\dot{z}=\mathrm{d}z/\mathrm{d}t$, $t\in\mathbb{C}$, and $V_{\mathscr{p}}(z)$ is the potential (\ref{def:Homotopy_potential}) evaluated at complex argument
\begin{equation}\label{def:complex_potential}
    V_{\mathscr{p}}(z) = \lambda(z^2-\eta^2)^2\velg{(1-\mathscr{p})z^2 + \xi^2\mathscr{p}}.
\end{equation}
As the potential (\ref{def:Homotopy_potential}) is a polynomial, it extends uniquely to an entire function of $z$, so no new definition is involved.

Exploiting the $Z_2$ symmetry of the system, and following the treatment of the double well \citep{Richard1981a,Tanizaki2014,Dersy2026b}, the solution to Eq.~(\ref{def:complex_EOM}) can be written as 
\begin{equation}\label{def:complex_solution}
    z(t) = \pm \sqrt{\frac{3a_3}{12\wp(t-t_0,g_2,g_3)-a_2}},
\end{equation}
with the parameters $a_2,a_3,g_2,g_3$ related to $\eta,\xi,\lambda,\mathscr{p}$ and $C$ as derived in Appendix \ref{Appendix_B}.

Denote $e_1,e_2,e_3$ the roots of the cubic polynomial
\begin{equation}
    4\wp^3 - g_2 \wp - g_3 = 4(\wp-e_1)(\wp-e_2)(\wp-e_3),
\end{equation}
the following identity between the Weierstrass and Jacobi elliptic $\mathrm{sn}$ functions holds \citep{Abramowitz2013}:
\begin{equation}\label{eq:jacobi_identities}
    \wp(t) = e_3 + \frac{e_1-e_3}{\text{sn}^2(\sqrt{e_1-e_3}t,m)}, \hmez m = \frac{e_2 - e_3}{e_1-e_3}, \hmez e_1\geq e_2\geq e_3.
\end{equation}
Using relation (\ref{eq:jacobi_identities}) in the solution (\ref{def:complex_solution}) results in
\begin{equation}\label{eq:jacobi_sn}
    z(t) = \pm\frac{\sqrt{3a_3}\,\text{sn}(v,m)}{\sqrt{12(e_1-e_3) + (12e_3-a_2)\,\text{sn}^2(v,m)}}, \qquad v = \sqrt{e_1-e_3}\,(t-t_0),
\end{equation}
which has an identical form as the instanton solution (\ref{def:instanton_solution}) where the hyperbolic tangent is replaced by an elliptic generalization.

The solution (\ref{def:complex_solution}) contains two integration constants corresponding to the instanton center $t_0$ and the complex energy $C$. In complex time (\ref{def:complex_time}) we make the substitution 
\begin{equation}\label{eq:jacobi_representation}
    z(\tau) = \pm \sqrt{\frac{3a_3}{12\wp(e^{-\mathrm{i}\alpha}(\tau-\tau_0),g_2,g_3)-a_2}}.
\end{equation}

Imposing the boundary conditions $\dot{z}(\pm\eta) = 0$  uniquely fixes $C=0$.  Algebraically, setting $C=0$ has a profound effect on the underlying elliptic function. The discriminant $\Delta = g_2^3 - 27g_3^2$ of the associated cubic equation (\ref{weierstrass}) vanishes ($\Delta = 0$), meaning that two roots coincide. Indeed, at $C=0$ the cubic of Appendix~\ref{Appendix_B} factorizes in the variable $y=1/z^2$ as $a_3y^3+a_2y^2+a_1y+a_0=-8\Lambda(1-\eta^2y)^2(y+b^2/\eta^2)$: the double zeros of $V_\mathscr{p}$ at the vacua $z=\pm\eta$ give the double root $e_2=e_3=-\omega^2/12$, and the imaginary turning points give $e_1=\omega^2/6$. Under the descending convention $e_1 \geq e_2 \geq e_3$, this fixes the parameter in the Jacobi function representation (\ref{eq:jacobi_sn})  to $m = 0$. This is useful as the following identity holds 
\begin{equation}
    \mathrm{sn}(v,m=0) = \sin(v).
\end{equation}
Since $e_1-e_3=\omega^2/4$, $v=\omega(t-t_0)/2$, which along the rotated time axis of Eq.~(\ref{eq:jacobi_representation}) with $\tau_0=\tau_c$ is $v=-\mathrm{i}\tilde u$. If $v$ is purely imaginary ($\alpha=\pi/2$), the solution loses all periodicity and we recover the Euclidean instanton solution (\ref{def:instanton_solution}). On the other hand if $v \in \mathbb{R}$ ($\alpha=0$), we get a purely periodic motion corresponding to the real time periodic instanton. Equivalently, in the language of elliptic functions, one period of $\wp$ stretches to infinity while the other remains finite, so the doubly-periodic lattice degenerates into a singly-periodic one. The other possible degeneration, $e_1=e_2$ ($m=1$, $\mathrm{sn}(u,1)=\tanh(u)$), occurs for $\mathscr{p}<\mathscr{p}_c$ at $C$ equal to the barrier-top (sphaleron) energy $V_\mathscr{p}(z_*)=4\Lambda\eta^4(1+b^2)^3/(27b^4)$, $z_*^2=\eta^2(b^2-2)/(3b^2)$. It describes real-time motion asymptotic to the barrier tops rather than tunneling, and is excluded by the instanton boundary conditions. For $\mathscr{p}>\mathscr{p}_c$ the barrier top is at $z=0$, with energy $V_\mathscr{p}(0)=\Lambda\eta^4$ at which $a_3=0$: the first integral (\ref{w_transform}) then has a double root at $w=0$ and is solved by elementary functions, so this case is not covered by Eq.~(\ref{def:complex_solution}). The two degenerations $m=0$ and $m=1$ thus mark the two ends of the finite-energy (periodic-instanton) family, the instanton at $C=0$ and the sphaleron at the barrier top.

The classical turning points of (\ref{def:complex_potential}) are given by double roots at 
\begin{equation}
    z = \pm \eta,
\end{equation}
and simple roots at 
\begin{equation}\label{eq:imag_turning_points}
    z = \pm \mathrm{i} \frac{\eta}{b}.
\end{equation}
We see that the imaginary turning points (\ref{eq:imag_turning_points}) are  exactly the bounds of the analytically continued instanton in real time (\ref{eq:bound}).  Using (\ref{eq:imag_velocity}) and the elliptic function structure, we conclude that the singularities encountered in the analytically continued double well instanton solution $X^{\alpha =0}_{\mathscr{p}= 1}$ are caused by the imaginary turning points (\ref{eq:imag_turning_points}) moving to infinity. An instanton connects two turning points and there exists a moment at which the classical solution connects the two infinities making the solution diverge. On the other hand, the analytically continued instanton for $\alpha \neq 0$ connects the two real turning points which are independent of $b$.

The mechanism can be made explicit. At $C=0$, real-time motion along the imaginary axis, $z=\mathrm{i}y$, obeys $\dot y^2=2V_\mathscr{p}(\mathrm{i}y)$ with $V_\mathscr{p}(\mathrm{i}y)=\lambda(y^2+\eta^2)^2\,[\xi^2\mathscr{p}-(1-\mathscr{p})y^2]$. In the double-well limit $\mathscr{p}=1$ the bracket is constant, $\dot y=\sqrt{2\lambda}\,\xi\,(y^2+\eta^2)$, and $y=\eta\tan(\omega t/2)$ reaches infinity in the finite time $\omega t/2=\pi/2$: this is the real-time comb, Eq.~(\ref{eq:realtime}) at $b=0$. That the quartic potential permits escape to infinity in finite time was noted in the Euclidean setting by \citet{Dersy2026b}. Complex paths that run to infinity and back in finite time also underlie the singularity crossings found by \citet{Feldbrugge2023} for the Gaussian barrier. For $\mathscr{p}<1$ the bracket changes sign at $\abs{y}=\eta/b$, so the simple roots (\ref{eq:imag_turning_points}) act as reflecting turning points. Then, the escape is forbidden, and the orbit is the bounded periodic solution (\ref{eq:realtime})--(\ref{eq:bound}).

The classification of Sec.~\ref{sec:continuation} acquires a simple
geometric form in the $\tilde{u}$-plane. The singular points of the solution, which are poles of $z^2$ (square-root branch points of $z$) and which we call poles for brevity, lie at
\begin{equation}
\tilde u_{\rm pole}=\pm\,\theta+\mathrm{i}\pi\Bigl(k+\tfrac12\Bigr),
\qquad k\in\mathbb{Z},\qquad\theta=\mathrm{arsinh}(b),
\label{eq:poles}
\end{equation}
two vertical lines at $\mathrm{Re}(\tilde{u})=\pm\theta$, with poles spaced
$\pi$ apart along each. A Wick-rotation ray
$\tilde{u}(u)=u(\sin\alpha+\mathrm{i}\cos\alpha)$ with $\alpha\neq0$ crosses each line
exactly once, which reproduces the count of Sec.~\ref{sec:continuation} and passes through a pole precisely when the
crossing height equals $\pi(k+\tfrac12)$, i.e.\ when
$\kappa(\alpha,b)=\theta/(\pi\tan\alpha)$ is a half-odd integer. At $\alpha=0$ the ray coincides with the line
$\mathrm{Re}(\tilde{u})=0$, which is a pole line if and only if $\theta=0$,
i.e.\ $b=0$ which is the Dirac comb.

Algebraically, the Weierstrass elliptic function is a doubly periodic meromorphic function. The double period creates a lattice in the complex plane. Thus, there exists a lattice of points for which $12\wp(z_i) = a_2$, making the solution $z(t)$ diverge. Using the complex Wick rotation, the time axis passes through at most two singularities. However as we go to real time, the time axis aligns with the direction of the single finite period that survives the $C=0$ degeneration of the $\wp$ function.

\begin{figure}[h!]
    \centering
    \includegraphics[width=0.7\linewidth]{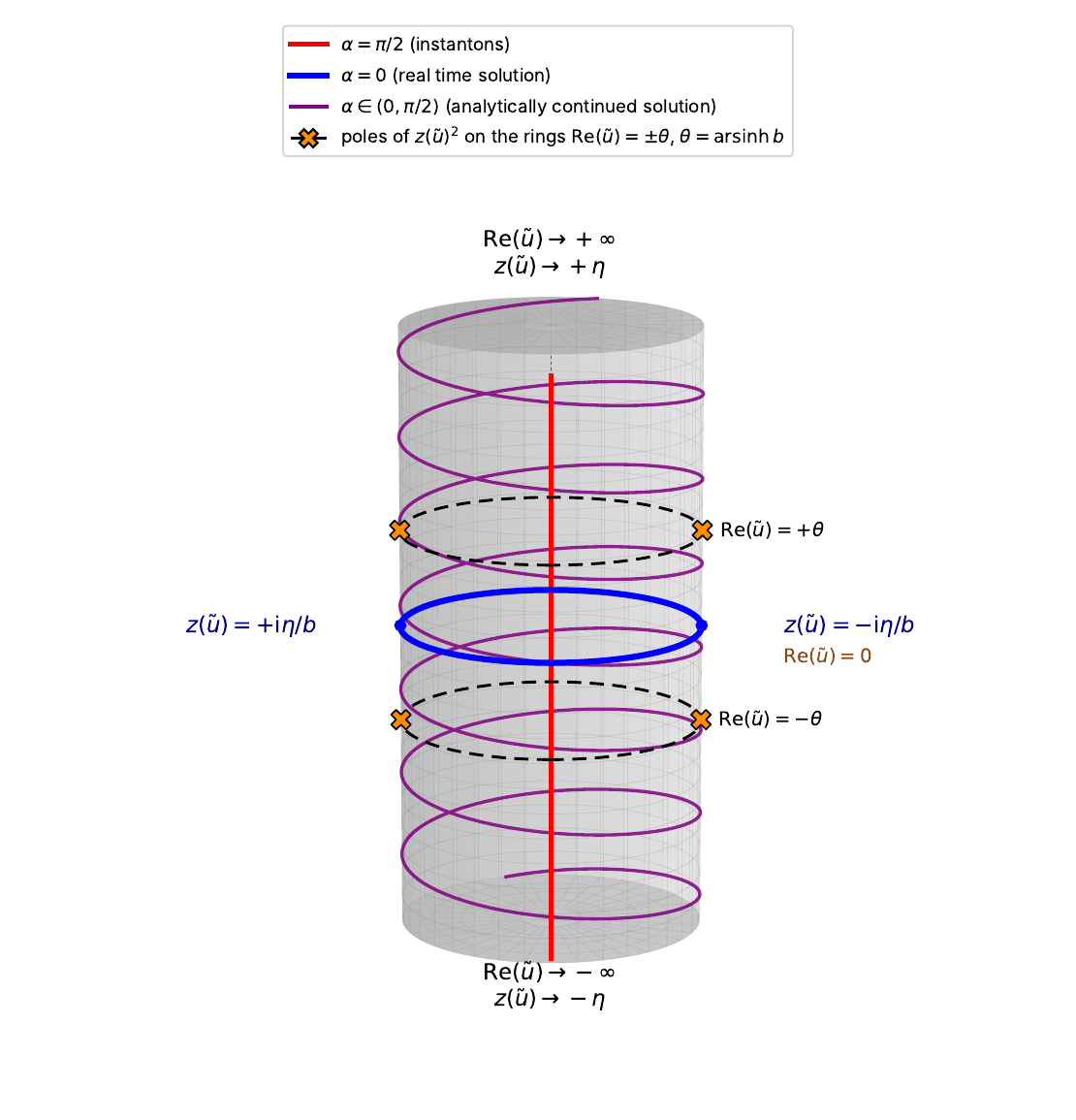}
    \caption{The pinched ($C=0$) genus-1 surface, represented as a cylinder in the complexified argument
$\tilde u = u\,e^{i(\pi/2-\alpha)}$; at $C=0$, $z(\tilde u)$ is the continued instanton $X(\tilde u)$ of Eq.~(\ref{def:complex_time_trajectory}). The axial direction, $\mathrm{Re}(\tilde u)$, is unbounded and
corresponds to the Euclidean solution ($\alpha=\pi/2$, red), which interpolates monotonically between
the classical vacua $z(\tilde u)\to\pm\eta$ as $\mathrm{Re}(\tilde u)\to\pm\infty$. The angular
direction, $\mathrm{Im}(\tilde u)$, is periodic and corresponds to the real-time solution
($\alpha=0$, blue), a closed loop on which the solution oscillates between its bounded extrema
$z(\tilde u)=\pm\mathrm{i}\eta/b$ (Eq.~(\ref{eq:bound})), the imaginary turning points of Eq.~(\ref{eq:imag_turning_points}) realized as the
left/right extremes of the loop. The purple curve shows a generic Wick-rotated trajectory,
$\alpha\in(0,\pi/2)$, which winds around the cylinder while drifting along its axis, interpolating
between the two degenerate limits. The dashed rings at $\mathrm{Re}(\tilde u)=\pm\theta$, $\theta=\mathrm{arsinh}(b)$, carry the poles (\ref{eq:poles}) of $z(\tilde u)^2$ (crosses), two per ring in each fundamental cell: every Wick-rotated trajectory crosses each ring exactly once and hits a pole only when Eq.~(\ref{eq:compatibility}) holds, while the real-time loop meets them only for $b=0$.}
    \label{fig:cylinder}
\end{figure}

Topologically, a unit cell of the periodic lattice is homeomorphic to a torus. The pinching at $C=0$ degenerates the torus into a cylinder, periodic along the surviving finite direction and unbounded
along the other (see Fig. \ref{fig:cylinder}). The poles of the resulting solution form two discrete, evenly spaced rows
along this single periodic direction (Eq.~(\ref{eq:poles})), rather than filling the surface. A generic Wick-rotation ray at
angle $\alpha$ crosses each of these rows exactly once, and hence passes through at most two poles, before escaping along the unbounded direction, which
is why never more than two singularities occur for $\alpha\neq0$. Only the special ray $\alpha=0$,
aligned with the surviving periodic direction, coincides with a row of poles only at the degenerate point $b=0$, at which the
entire row of poles is crossed simultaneously, producing the infinite Dirac comb of Cherman and Ünsal
as the one non-generic case.

\subsection{Action of the continued instanton}\label{sec:action}
The pole structure also determines whether the action of the continued instanton depends on $\alpha$. For the double well, Cherman and Ünsal showed that it does not \citep{Cherman2014}. By the BPS equation (\ref{def:BPS_equation}), the action of the continued solution (\ref{def:complex_time_trajectory}) is
\begin{equation}\label{eq:Salpha}
S(\alpha)=\frac{\omega}{2}\int_{\Gamma_\alpha}X'(\tilde u)^2\,\mathrm{d}\tilde u,
\qquad \Gamma_\alpha:\ \tilde u=u\,e^{\mathrm{i}(\pi/2-\alpha)},\ u\in\mathbb{R},
\end{equation}
which reduces to $S_{\mathrm{E}}^0$ of Eq.~(\ref{eq:action}) at $\alpha=\pi/2$. The integrand
\begin{equation}\label{eq:Xp2}
X'(\tilde u)^2=\frac{\eta^2(1+b^2)^2\cosh^2\tilde u}{(\cosh^2\tilde u+b^2)^3}
\end{equation}
is meromorphic, since squaring removes the branch cut of $X$ itself, with triple poles exactly at the points (\ref{eq:poles}), and it decays as $|\mathrm{Re}\,\tilde u|\to\infty$. Rotating $\Gamma_\alpha$ is therefore a contour deformation: $S(\alpha)$ is constant as long as $\Gamma_\alpha$ crosses no pole, and changes by $2\pi\mathrm{i}$ times the enclosed residues when it does. The residue at every pole of the row $\mathrm{Re}\,\tilde u=\theta$ is (Appendix~\ref{app:residue})
\begin{equation}\label{eq:residue}
\operatorname*{Res}_{\tilde u=\theta+\mathrm{i}\pi(k+\frac12)}X'(\tilde u)^2
=-\frac{\eta^2(1+4b^2)}{16\,b^3\sqrt{1+b^2}},
\end{equation}
independent of $k$ by the $\mathrm{i}\pi$-periodicity of (\ref{eq:Xp2}), and opposite on the row $\mathrm{Re}\,\tilde u=-\theta$ because (\ref{eq:Xp2}) is even. As $\alpha$ decreases, $\Gamma_\alpha$ crosses the pair $\pm(\theta+\mathrm{i}\pi(k+\frac12))$ precisely when $\kappa(\alpha,b)=k+\frac12$, i.e.\ at the singular angles $\alpha^{\mathrm{singular}}_{k}(b)=\arctan\!\big[\theta/(\pi(k+\tfrac12))\big]$ of Eq.~(\ref{eq:compatibility}). Each crossing changes the action by
\begin{equation}\label{eq:jump}
\Delta S=-4\pi\mathrm{i}\,\frac{\omega}{2}\operatorname*{Res}X'^2
=\mathrm{i}\,\frac{\pi\eta^2\omega\,(1+4b^2)}{8b^3\sqrt{1+b^2}}=-\mathrm{i}\,S_{\text{period}},
\end{equation}
where $S_{\text{period}}$ is the Minkowski action (\ref{eq:Speriod}) of one period of the real-time orbit. Hence
\begin{equation}\label{eq:staircase}
S(\alpha)=S_{\mathrm{E}}^0-\mathrm{i}\,N(\alpha)\,S_{\text{period}},\qquad
N(\alpha)=\#\{k\ge0:\ \alpha^{\mathrm{singular}}_{k}(b)>\alpha\}.
\end{equation}

The result can also be read in the complex $X$-plane. Since $\frac{\omega}{2}X'(\tilde u)=\sqrt{2V_\mathscr{p}(X)}$ holds identically, Eq.~(\ref{eq:Salpha}) is $S(\alpha)=\int\sqrt{2V_\mathscr{p}(X)}\,\mathrm{d}X$ along the image of $\Gamma_\alpha$, which is the form in which Cherman and Ünsal proved the $\alpha$-independence for the double well \citep{Cherman2014}: there $\sqrt{2V}=\sqrt{2\Lambda}(\eta^2-X^2)$ is a polynomial and the integral depends only on the end points $\mp\eta$. For $b>0$, $\sqrt{2V_\mathscr{p}}=\sqrt{2\Lambda}\,(\eta^2-X^2)\sqrt{1+b^2X^2/\eta^2}$ has branch points at the imaginary turning points (\ref{eq:imag_turning_points}), outside the cut joining them it is single valued, and at infinity
\begin{equation}\label{eq:sqrtV_infinity}
\sqrt{2V_\mathscr{p}(X)}=\sqrt{2\Lambda}\,\frac{b}{\eta}\Bigl[-X^3+\eta^2\Bigl(1-\frac{1}{2b^2}\Bigr)X+\frac{\eta^4(1+4b^2)}{8b^4}\,\frac{1}{X}+O(X^{-3})\Bigr].
\end{equation}
Near a pole, $X\propto(\tilde u-\tilde u_{\rm pole})^{-1/2}$, so a circuit of $\tilde u$ around each of the two poles crossed at a singular angle maps to half a turn of $X$ around infinity. The pair adds one loop enclosing the cut, and the jump is $2\pi\mathrm{i}$ times the coefficient of $1/X$ in (\ref{eq:sqrtV_infinity}),
\begin{equation}\label{eq:jump_residue}
\Delta S=2\pi\mathrm{i}\,\sqrt{2\Lambda}\,\eta^3\,\frac{1+4b^2}{8b^3}=-\mathrm{i}\,S_{\text{period}},
\end{equation}
which with $\omega=2\sqrt{2\Lambda}\,\eta\sqrt{1+b^2}$ is identically Eq.~(\ref{eq:jump}). The residue at infinity vanishes for the double well, which is the $\alpha$-independence of Cherman and Ünsal; for $b\to0$ it grows like $b^{-3}$ while the singular angles $\alpha^{\mathrm{singular}}_{k}\to0$, so that at any fixed $\alpha>0$ the action stays constant, in agreement with Sec.~\ref{sec:continuation}.

\begin{figure}[h!]
    \centering
    \includegraphics[width=0.62\linewidth]{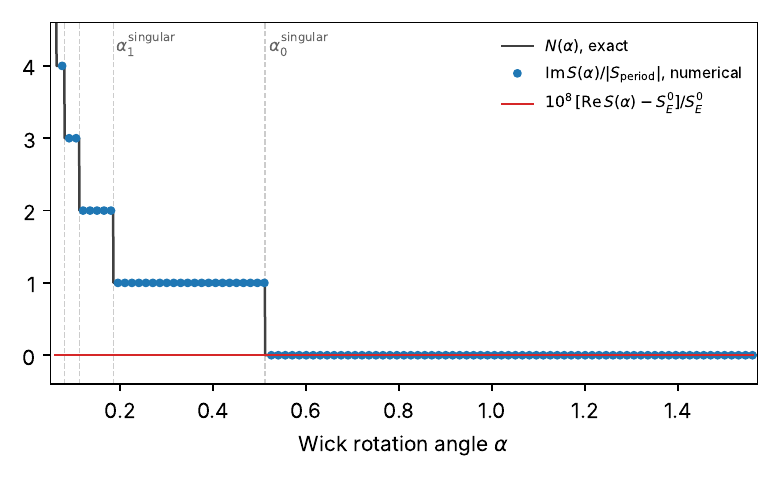}
    \caption{Action of the continued instanton, Eq.~(\ref{eq:Salpha}), as a function of the Wick rotation angle for $b=1$, $\eta=1$. Points: direct numerical integration of (\ref{eq:Salpha}), imaginary part in units of $|S_{\text{period}}|$. Line: the staircase $N(\alpha)$ of Eq.~(\ref{eq:staircase}). Red: the relative deviation of the real part from $S_{\mathrm{E}}^0$, magnified by $10^{8}$. Dashed lines mark the singular angles $\alpha^{\mathrm{singular}}_{k}$.}
    \label{fig:staircase}
\end{figure}

Three consequences follow. First, the real part of the action, which sets the tunneling suppression $|e^{-S}|=e^{-\mathrm{Re}\,S}$, is independent of $\alpha$ throughout the family. Second, the $\alpha$-independence of the full action found by Cherman and Ünsal is a special property of the double well: at $b=0$ the poles lie on $\mathrm{Re}\,\tilde u=0$ and are never crossed by $\Gamma_\alpha$ for $\alpha>0$, so that (\ref{eq:staircase}) reduces to $S=S_{\mathrm{E}}^0$. For $b>0$ the imaginary part is a staircase whose steps accumulate as $\alpha\to0$, where the ray turns into the periodic direction of the cylinder of Fig.~\ref{fig:cylinder}. Third, the singular potentials of Sec.~\ref{sec:continuation} acquire a meaning beyond the divergence of the solution: they are the points at which the continued instanton winds once more around the surviving cycle of the pinched curve, the $C=0$ counterpart of the winding number of the finite-$T$ saddles of \citet{Dersy2026a}, and picks up one period of the real-time orbit. These jumps are singularity crossings in the sense of \citet{Feldbrugge2023,Feldbrugge2025}, at which the action of an analytically continued complex path acquires a discrete branch-cut correction. Eq.~(\ref{eq:staircase}) provides an exact, closed-form instance in which this correction is identified with the real-time period action. A direct numerical integration of (\ref{eq:Salpha}) confirms (\ref{eq:staircase}): for $b=1$ and $b=\frac12$ the real part is constant to ten significant digits, and the imaginary part equals $N(\alpha)$ times the predicted jump to nine (Fig.~\ref{fig:staircase}).

\section{Discussion}\label{sec:discussion}
The singularities encountered in the analytically continued instanton solution of the double-well potential constructed by Cherman and Ünsal are not directly caused by the real-time limit, but by the turning points of a more general system moving to infinity. We constructed a family of potentials that are everywhere regular and bounded in the real-time limit for any $\mathscr{p}\in(0,1)$, and showed that for every $\alpha\in(0,\pi/2)$ there exists a countably infinite, closed-form family of potential parameters $\{\mathscr{p}^{\mathrm{singular}}_{k}(\alpha)\}_{k\geq0}$, each producing exactly two singularities on the rotated time axis
and never more. Regularity is therefore generic, and singular behaviour occurs only on a discrete, measure-zero set of potentials for any fixed $\alpha\neq0$. The general behaviour of the system, including the origin of both regimes, can be described by the Weierstrass elliptic function.

Section~\ref{sec:action} also bears on the Cherman-Ünsal conjecture recalled in the Introduction \citep{Cherman2014}. In the homotopy family only the real part of the action is $\alpha$-independent, while the imaginary part jumps by $-\mathrm{i} S_{\text{period}}$ at every singular angle, Eq.~(\ref{eq:staircase}). Any formulation of minimality within the family must therefore refer to $\mathrm{Re}\,S$. The double-well statement is recovered as the degenerate case without singularity crossings in the sense of \citet{Feldbrugge2023}.

The genus classification of Section \ref{sec:homotopy} also indicates a natural direction for extending this construction. Homotopies with $n \geq 5$ reduce to higher genus hyperelliptic systems, for which no elementary closed-form instanton solution exists, but which may still be tractable via the uniform WKB \citep{Dunne2014a} and Bender-Wu \citep{Sulejmanpasic2016} methods planned for future work. Because such homotopies necessarily skip intermediate well-counts (Section \ref{sec:homotopy}), they would connect potentials of non-consecutive well-number directly which is a qualitatively different construction from the one studied here, and one where the interplay between genus, singularity structure, and resurgent behaviour remains open.

We emphasize that the classification given here is purely classical. Whether the singular and periodic saddles identified above actually contribute to the real-time path integral is a question of Picard-Lefschetz intersection numbers that we have not computed, and Section~\ref{sec:weierstrass} should be read as setting up that problem rather than solving it. In Euclidean time at finite period, the analogous question has recently been settled for the double well: only real saddles contribute to the partition function, while the complex ones govern its Stokes structure \citep{Dersy2026a}. The real-time counterpart for the saddles identified here remains open. The bounded, purely imaginary saddle of Eqs. (\ref{eq:bound})-(\ref{eq:imag_velocity}) is a close relative of the ghost instantons of Başar, Dunne and Ünsal \citep{Basar2013}, and the finite-energy periodic solutions of Section~\ref{sec:continuation} sit in the same family as the periodic instantons of Liang and Müller-Kirsten \citep{Liang1992} and of Khlebnikov, Rubakov and Tinyakov \citep{Khlebnikov1991}. A proper accounting of our results within that literature, and an explicit thimble analysis of the oscillatory, unsuppressed weight $e^{\mathrm{i}S_{\text{period}}}$, are natural next steps. That this programme can be completed is demonstrated by \citet{Ai2019a}, who recovered the full one-loop Callan-Coleman decay rate of the false vacuum from the real-time path integral over Lefschetz thimbles.

\section*{Acknowledgements}
We are deeply grateful to Klaus Bering Larsen for his valuable insights and feedback on previous versions of this manuscript. V.L.\ acknowledges support from the Faculty of Mathematics and Physics, Charles University, under the Student Faculty Grant (SFG 2025). T.S. and Š.K. acknowledge support from the Ministry of Education, Youth and Sports of the Czech Republic under project No.~LM2023040 "Research Infrastructure for Experiments at CERN" (CERN-CZ). Š.K. was supported by the project Quantum materials for applications in sustainable technologies (QM4ST), funded as Project No. \verb|CZ.02.01.01/00/22_008/0004572| by Programme Johannes Amos Comenius, call Excellent Research. During the preparation of this manuscript, the authors used an AI language model (Anthropic; model versions Claude Opus 5.5 and Claude
Fable 5) to review drafts for mathematical and logical errors, to suggest arguments and formulations, to perform numerical checks, to search for and verify references, to draft initial outlines of specific paragraphs, and to refine the overall text style. The model was directed through iterative prompts containing the manuscript text and specific questions. Following this process, the authors thoroughly reviewed and rewrote the content and checked all derivations and references. The conceptual design of the work was developed by the authors, who take full responsibility for the final manuscript.

\section*{Data availability}
All figures are plots of the closed-form expressions given in the text, except Fig.~\ref{fig:staircase}, which shows a direct numerical integration of Eq.~(\ref{eq:Salpha}) that can be reproduced from the expressions given in the text. No other data were created or analyzed in this study.

\appendix 
\setcounter{figure}{0} 

\section{Derivation of the solution to the complexified equations of motion}\label{Appendix_B}

To solve the following differential equation
\begin{equation}
    (\dot{z})^2 = 2\velz{C-\lambda(z^2-\eta^2)^2 \velg{(1-\mathscr{p})z^2  + \xi^2 \mathscr{p}}},
\end{equation}
we reduce the sixth order polynomial using the substitution $w = z^2$ resulting in
\begin{equation}\label{w_transform}
    (\dot{w})^2 = 8w \velz{C-\lambda(w-\eta^2)^2 \velg{(1-\mathscr{p})w + \xi^2 \mathscr{p}}}.
\end{equation}
This is a quartic polynomial. We now reduce it to a third order polynomial using the substitution
\begin{equation}
    y = \frac{1}{w} \implies (\dot{w})^2 = \frac{(\dot{y})^2}{y^4}.
\end{equation}
The resulting differential equation for $y$ is
\begin{align}\label{rovnice_pro_y}
    (\dot{y})^2 = 8y^3(C-\lambda\eta^4\xi^2 \mathscr{p}) + 8y^2(2\lambda \eta^2\xi^2 \mathscr{p}-\lambda \eta^4 (1-\mathscr{p} )) + \\
    8y(2\lambda\eta^2(1-\mathscr{p})-\lambda\xi^2\mathscr{p}) - 8\lambda(1-\mathscr{p}).
\end{align}
which can be rewritten as
\begin{equation}
    (\dot{y})^2 = a_3y^3 + a_2 y^2+a_1y+a_0,
\end{equation}
where
\begin{gather}\label{weierstrass a}
     a_0 = -8\lambda(1-\mathscr{p}) = -8\Lambda\frac{b^2}{\eta^2}, \\
     a_1 = 8(2\lambda\eta^2(1-\mathscr{p})-\lambda\xi^2\mathscr{p}) = 8\Lambda(2b^2-1),\\
     a_2 = 8(2\lambda \eta^2\xi^2\mathscr{p}-\lambda \eta^4 (1-\mathscr{p})) = 8\Lambda\eta^2(2-b^2),\\
     a_3 = 8(C-\lambda\eta^4\xi^2\mathscr{p}) = 8(C-\Lambda\eta^4),
\end{gather}
with $\Lambda=\lambda\xi^2\mathscr{p}$ and $b^2$ as in Eqs.~(\ref{def:b}) and (\ref{eq:oneparam}), using $\lambda(1-\mathscr{p})=\Lambda b^2/\eta^2$.

A linear shift
\begin{equation}
    y(t) = \frac{4}{a_3}  \wp(t) - \frac{a_2}{3a_3} \iff  \wp(t) = \frac{a_3}{4}y(t) + \frac{a_2}{12},
\end{equation}
 eliminates the quadratic term. Then we get the defining equation of the Weierstrass elliptic function \citep{Abramowitz2013}
\begin{equation}\label{weierstrass}
    \dot{\wp}^2(t-t_0,g_2,g_3) = 4 \wp^3 - g_2  \wp - g_3,
\end{equation}
where
\begin{gather}\label{weierstrass g}
    g_2 = \frac{1}{12}(a_2^2 - 3a_1a_3),\\
    g_3 = \frac{1}{432} (9a_1a_2a_3 - 2a_2^3 - 27a_0 a_3^2).
\end{gather}

\section{Visualizations of the analytically continued instanton solution}\label{AppendixB}

In this Appendix, we present the full-page comprehensive numerical plots corresponding to the linear homotopy deformation discussed in Sec.~\ref{sec:continuation}.  Due to the high density of the data points, these visualization profiles are provided as standalone figures to ensure clarity and readability.

\begin{figure*}
    \centering
    \includegraphics[width=0.8\linewidth]{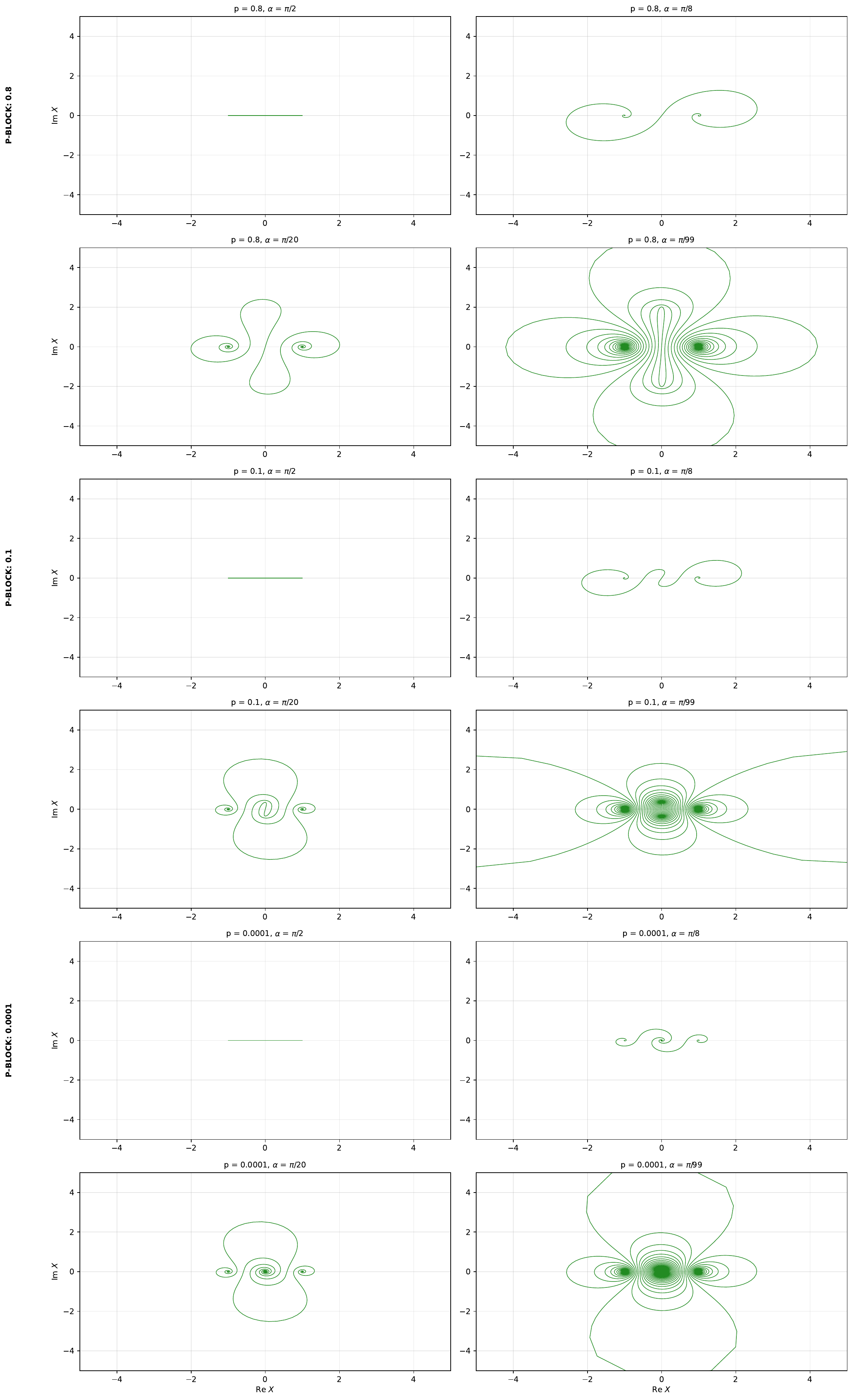}
    \caption{Analytically continued classical trajectories of the homotopy potential in the complex plane as a function of the angle $\alpha$ and parameter $\mathscr{p}$.  As $\alpha$ decreases toward the real-time limit, the trajectories evolve into complex, spiral-like shapes. Parameters used: $\lambda = 0.1, \eta = 1, \xi = 1$.}
    \label{Huge_phase_space}
\end{figure*}

\begin{figure*}
    \centering
    \includegraphics[width=0.8\linewidth]{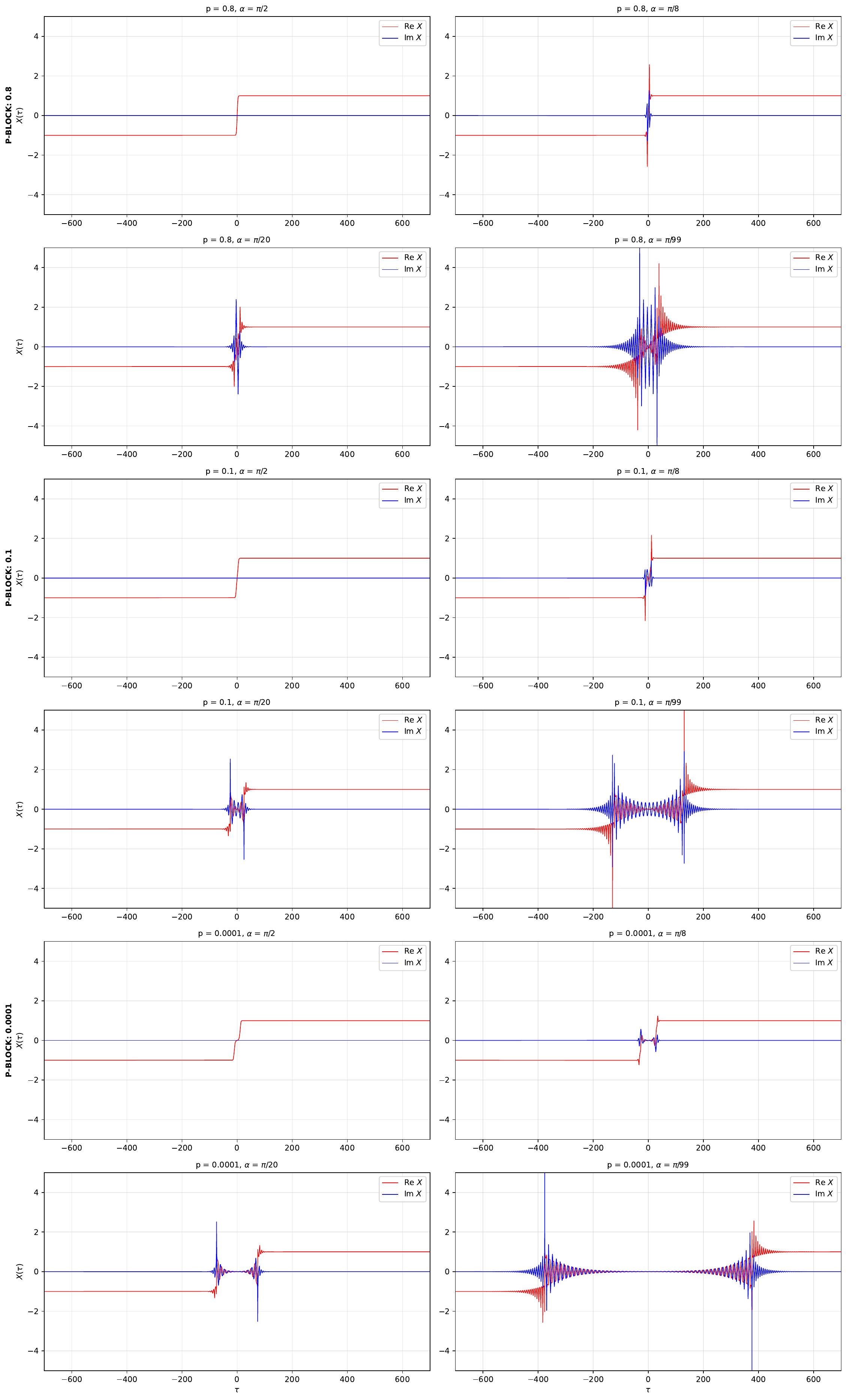}
    \caption{Evolution of the real ($\text{Re} X^\alpha_\mathscr{p}$) and imaginary ($\text{Im} X^\alpha_\mathscr{p}$) components of the analytically continued classical trajectory across various angles $\alpha$ and parameters $\mathscr{p}$.  As $\alpha$ decreases, the imaginary component $\text{Im} X^\alpha_\mathscr{p}$ emerges as a series of oscillations centered around the tunneling event. This reflects the spiral-like geometry of the complexified path and the transition toward the oscillatory features of the real-time limit. In the triple-well regime (lower panels), the solution clearly retains a double-transition profile. Parameters used: $\lambda = 0.1, \eta = 1, \xi = 1, \tau_c = 0$.}
    \label{Re-Im_evolution}
\end{figure*}

\newpage
\section{Residue of \texorpdfstring{$X'(\tilde u)^2$}{X'(u)^2}}\label{app:residue}
Let $\tilde u_p=\theta+\mathrm{i}\pi/2$, $w=\tilde u-\tilde u_p$ and $B=\sqrt{1+b^2}$. Since $\cosh\tilde u_p=\mathrm{i} b$ and $\sinh\tilde u_p=\mathrm{i} B$,
\begin{equation}
\cosh\tilde u=\mathrm{i}\,h(w),\qquad
h(w)=b\cosh w+B\sinh w=b+Bw+\tfrac{b}{2}w^2+\tfrac{B}{6}w^3+O(w^4).
\end{equation}
Then $\cosh^2\tilde u+b^2=b^2-h^2=-(h-b)(h+b)=-w\,Q(w)$ with
\begin{equation}
Q(w)=2bB+(1+2b^2)\,w+\tfrac43\,bB\,w^2+O(w^3),
\end{equation}
and Eq.~(\ref{eq:Xp2}) becomes $X'^2=\eta^2B^4h^2/(w^3Q^3)$. The residue is $\eta^2B^4$ times the coefficient of $w^2$ in $h^2/Q^3$. Writing $Q=2bB\,(1+q_1w+q_2w^2)$ with $q_1=(1+2b^2)/(2bB)$ and $q_2=\frac23$, and $h^2=b^2+2bB\,w+(1+2b^2)\,w^2+O(w^3)$, this coefficient is
\begin{equation}
\frac{1}{(2bB)^3}\Big[b^2\big(6q_1^2-3q_2\big)-6bB\,q_1+1+2b^2\Big]
=-\frac{1}{8b^3B^3}\,\frac{1+4b^2}{2B^2},
\end{equation}
which gives Eq.~(\ref{eq:residue}). The other poles follow from the $\mathrm{i}\pi$-periodicity and the parity of (\ref{eq:Xp2}).

\printbibliography
\end{document}